\documentclass[twocolumn,showpacs,preprintnumbers,amsmath,amssymb]{revtex4}

\usepackage{graphicx}
\usepackage{epsfig}

\begin{document}
\preprint{APS/123-QED}

\title{Testing the principle of equivalence with Planck surveyor}
\author{L.A.~Popa\footnote{Also at the Institute of Space Sciences,
Bucharest, Romania}, C.~Burigana and N.~Mandolesi} 
\affiliation{IASF/CNR, 
Istituto di Astrofisica Spaziale e Fisica Cosmica,
Sezione di Bologna,\\
Consiglio Nazionale delle Ricerche, Via Gobetti 101, I-40129 Bologna,
Italy }
\date{\today}
\begin{abstract}
We consider the effect of the violation of the
equivalence principle (VEP) by the massive neutrino component on the
Cosmic Microwave Background angular power specrum. 
We show that in the
presence of   
adiabatic and isocurvature primordial density perturbations 
the {\sc Planck} surveyor can place limits 
on the maximal VEP by the massive 
neutrino component at the level of 
${\rm few}\times 10^{-5}$, valid in the general
relativity, for the case in which the gravity is the single source of VEP. 
This work has been performed within the framework of the 
{\sc Planck}/LFI activities.

\end{abstract}

\maketitle

The atmospheric neutrino
experiments \cite{Fukuda98,Ambrosio98} 
provide strong evidence for neutrino
oscillations implying  a non-zero neutrino mass
with a lower limit $\simeq 0.04-0.08$ eV.
The latest measurements of the contribution of double beta decay
to the neutrino mass matrix \cite{Kla2001} place
an upper limit on the neutrino mass of $m_{\nu} \leq 0.26$eV.
The direct implication of these results is the
non-negligible contribution of the hot dark matter (HDM)
to  the total mass density of the universe and the existence of
three massive neutrino flavors
(i.e. a density parameter $\Omega_{\nu}h^2
\approx\sum_{i=1}^{3}m_i/93$eV,
where $h=H_0/100$ Km s$^{-1}$ Mpc$^{-1}$ is the dimensionless Hubble
constant). The introduction of a HDM component in the form
of massive neutrinos with the total mass in eV range
is also required
for consistency between the Cosmic Microwave Background (CMB) anisotropy
at small scales with the Large Scale Structure (LLS) of the universe
derived from galaxy surveys \cite{Scott94, White95, Primack95,
Gawiser98}.

The most popular mechanisms examined as possible solutions to
the neutrino oscillation problem
are the mass-mixing vacuum  and the matter induced
(MSW) oscillations  \cite{MSW},
having as basic underlying hypothesis
the existence of  non-degenerated neutrino masses
and the non equivalence of neutrino weak and mass eigenstates.

An alternative mechanism to these
approaches is the neutrino oscillations due to the Violation of the
Equivalence Principle (VEP) that 
can occur if the week
equivalence principle is not satisfied, i.e. the coupling of neutrinos
with the gravitational field is non-universal and the flavor eigenstates
are not identical to the states that couple to gravity
\cite{Gas88, Gas89, Har91, Har96}.
Such mechanism has been investigated by using the experimental results 
from Super-Kamiokande (SK) \cite{Gago2000, Casini2000} and
Sudbury Neutrino Observatory (SNO) \cite{Ray2002} collaborations.
Recent works have considered the possibility to test VEP by using
astrophysical neutrino sources \cite{Min95} as well as  the effect
of neutrino oscillations induced by VEP as an explanation 
of the pulsar kick velocity \cite{Horvat98}.

A deviation from the equivalence principle
can be parametrized by assuming that the
parametrized-post Newtonian (PPN)
parameters  for a given metric have
particle-dependent values \cite{Will93}.
In the conformal Newtonian gauge the perturbations
in the Robertson-Walker spacetime are characterized by two scalar
potentials $\Psi$ and $\Phi$ which appear in the line element
as \cite{Muk92}:
\begin{equation}
d \, s^2=a^2( \tau )
[-(1+2 \alpha \Psi) d \,\tau^2
+(1-2 \gamma \Phi) d \, x^i {\rm d} x_i] \, ,
\end{equation}
where $a(\tau)$ is the cosmic scale factor ($a_0$=1 today)
and $\tau$ is the conformal time.
$\alpha$ and $\gamma$ are theory dependent parameters:
in the general relativity $\alpha=\gamma=1$ while
in alternative theories of relativity satisfying
the equivalence principle
$\alpha$ and $\gamma$ need not to be
unity but their values are
universal (the same for all kinds of particles).
In eq.~(1) $\Psi$ plays the role of the
gravitational potential in the Newtonian limit
\cite{Ma95}.
Limits on the VEP for $\alpha=1$
have been obtained from the supernova SN1987A by
comparing the arrival time for photons and neutrinos
($|\gamma_{\gamma}-\gamma_{\nu}| <$ few $\times 10^{-3}$)
\cite{Longo88, Krauss88} and for neutrinos and
antineutrinos
($|\gamma_{ \nu_{e} }- \gamma_{ {\bar \nu}_e}| <  10^{-6}$)
\cite{LoSecco98, Pakvasa99}.

The CMB anisotropy  holds the
key for understanding the seeds of the cosmological structures 
and allows the measurement of the most important
cosmological parameters. The new generation of CMB experiments such as
MAP
\footnote{http://map.gsfc.nasa.gov} 
and
{\sc Planck}
\footnote{http://astro.estec.esa.nl/Planck}
will achieve enough precision to reveal 
the structure formation process up to arcminute angular scales.
The aim of this paper is to study the possibility to
test with {\sc Planck} 
the VEP by the massive neutrino component.

The evolution of the density fluctuations from the early universe
involves the integration of
coupled and linearized Boltzmann, Einstein and fluid equations
\cite{Ma95} 
for all the relevant
species (e.g., photons, baryons, cold dark matter, massless
and massive neutrinos).
The differences introduced in the gravitational
potential due to the VEP by the massive neutrino component can
generate metric perturbations that
affect the evolution  of the density
fluctuations of all the components,
leaving imprints on the CMB angular power spectrum.

A recent work \cite{Popa2000} analysed the possibility to detect  
neutrino oscillations with accurate CMB experiments.
In the presence of a large neutrino asymmetry
the mass-mixing vacuum neutrino oscillations
left detectable  imprints on the CMB anisotropy
and polarization power spectra.
For a neutrino flavor $\nu_i$ with energy $E_{\nu_i}$
and mass $m_{\nu_i} \leq 0.26$ eV,
the survival probability after the propagation through the distance $L$
in the expanding universe,
${\cal P}_{\nu_i, \nu_i}(E_{\nu_i},L)=1-\sin^2 \, 2\theta_0 \sin^2
( \pi L / \lambda )$,  
$\theta_0$ being the vacuum mixing angle, 
does not exhibit a dependence on the oscillation length $\lambda \sim
E_{\nu_i}$, as $L>> \lambda$ and $< \sin^2 (L/ \lambda)> \rightarrow 
1/2$. 

For the VEP induced oscillation,  the oscillation probability
is given by the same equation,  replacing 
$\theta_0$ with the gravitational mixing angle, $\theta_G$,
and with the oscillation length given by
$ \lambda \sim (\Psi \Delta \alpha_{\nu_{i}} E_{\nu_i})^{-1}$,
where $\Psi$ is the gravitational potential and
$\Delta \alpha_{\nu_i}=1-\alpha_i$
parametrizes the VEP by the neutrino flavor $\nu_{i}$ \cite{Har96, Ray2002}.
Also in the VEP case, 
$L>> \lambda$
for neutrinos with masses $m_{\nu} \leq 0.26$ eV and the
oscillation probability  is not sensitive
to the VEP parameter $\Delta \alpha_i$.
It follows that the mass-mixing vacuum oscillations can not be
distinguished from VEP induced oscillations
by using the CMB anisotropy.

On the other hand, the imprint of the clusterization process on the 
CMB anisotropy depends on the dynamics of the involved 
particles~\cite{Popa2002} and can be then sensitive to the VEP. 
We evaluate the effect of the VEP by the massive neutrinos
computing the CMB temperature
fluctuations generated in the non-linear
stages of the universe evolution when clusters and superclusters
of galaxies start to form.
We place  limits on the maximal VEP 
by the massive neutrinos in the general relativity,
valid for the case in which gravity is the only source of VEP.

In  \cite{Popa2002} we studied the CMB anisotropy
induced by the non-linear perturbations in
the massive neutrino density
associated with the non-linear
gravitational clustering process.
This process generates metric perturbations
leading to a decrease in the CMB anisotropy power spectrum
of amplitude $\Delta T/T \approx 10^{-6}$
for angular resolutions between $\sim 4$ and 20 arcminutes,
depending on the cluster mass scale $M_{clust}$ and the neutrino
fraction
$f_{\nu}=\Omega_{\nu}/(\Omega_b+\Omega_{cdm})$, where
$\Omega_{\nu}$, $\Omega_b$ and $\Omega_{cdm}$ are 
the energy density parameters for neutrinos, baryons and cold dark
matter particles, respectively. 
In the Newtonian limit, the neutrino gravitational clustering
can be described as a deviation
from the background by a potential $\Psi$
given by the Poisson equation:
\begin{equation}
{\nabla}^2 \Psi({\vec r},a) =4 \pi G a^2
\rho_m(a)\delta_m({\vec r},a) \, ,
\end{equation}
where ${\vec r}$ is the position 3-vector,
$\rho_m(a)$ is the
matter density
and $\delta_m({\vec r},a)$ is the matter density
fluctuation.
The equations governing the motion of each particle species
$i$ ($i$ = cold dark matter, baryons, neutrinos)
in the expanding universe
are given by \cite{Kates1991, Gleb1994}:
\begin{eqnarray}
\frac{{d \vec q_i}}{da}= -a\,H(a) \, \alpha_i{\vec \nabla} \Psi ,
\hspace{0.5cm}  \frac{ d{\vec r_i} } { da }
={\vec q_i}\, [a^3 \, H(a)]^{-1},
\end{eqnarray}
where 
${\vec q_i}$ is
the comoving momentum and $H(a)$ is the Hubble expansion
rate:
\begin{eqnarray}
H^2(a)=\frac{8 \pi G}{3}(\Omega_m/a^3+\Omega_r/a^4+
\Omega_{\Lambda}+\Omega_k/a^2) \, . \nonumber
\end{eqnarray}
Here $G$ is the gravitational constant,
$\Omega_m=\Omega_b+\Omega_{cdm}+\Omega_{\nu}$
is the matter energy density parameter, $\Omega_{r}$, 
$\Omega_{\Lambda}$ and $\Omega_k=1-\Omega_m-\Omega_{\Lambda}$
are the density parameters for 
radiation
(including the contribution from photons and relativistic neutrinos),
vacuum energy and the curvature of the universe.
The parameter $\alpha_i$ from the eq.~(3) accounts for
the coupling of each $i^{th}$ particle species with
the gravitational potential.
The Newtonian description given by eqs.~(2) and (3)
applies in the limit of the weak gravitational
field if, at each time
step, the size of the non-linear structures
is much smaller than the causal horizon
size (i.e., the background curvature is negligible). 

By using numerical simulations~\cite{Popa2002},
we evaluate the effect of the gravitational clustering
on the CMB anisotropy power spectrum  first
assuming $\alpha_i=1$ for all the
particles and then considering that
the equivalence principle is violated only by
massive neutrinos ($\alpha_{\nu} \ne 1$).

We consider a flat $\Lambda$CHDM cosmological model 
with $\Omega_m$=0.38, $\Omega_{\Lambda}$=0.62,
H$_0$=62~Km~s$^{-1}$~Mpc$^{-1}$, three massive neutrino flavors
giving a neutrino fraction $f_{\nu}$=0.06 
($m_\nu$=0.78eV, $\Omega_{\nu}$=0.022), 
and adopt a cluster mass $M_{clust}=$1.5$ \times 10^{15}M_{\odot}$. 
We assume a primordial power spectrum with adiabatic perturbations and 
a scalar spectral index $n^{adi}=0.98$ and neglect,
for simplicity, the contributions from the tensorial modes 
(gravitational waves) and the reionization effects.
This model is consistent with the LSS data
and the latest CMB anisotropy measurements,
allowing at the same time a pattern of neutrino masses
consistent with the results from neutrino oscillation
and double beta decay experiments.
Panel a) from Fig.~1 presents the CMB anisotropy power spectra
obtained from numerical simulations for different values
of $\Delta \alpha_{\nu} =1-\alpha_{\nu}$.
The VEP parameter $\Delta \alpha_{\nu} \neq 0$  induced metric
perturbations, leading to a decrease of the CMB temperature
anisotropy at small scales as $\Delta \alpha_{\nu}$ increases.

\begin{figure*}
\epsfig{figure=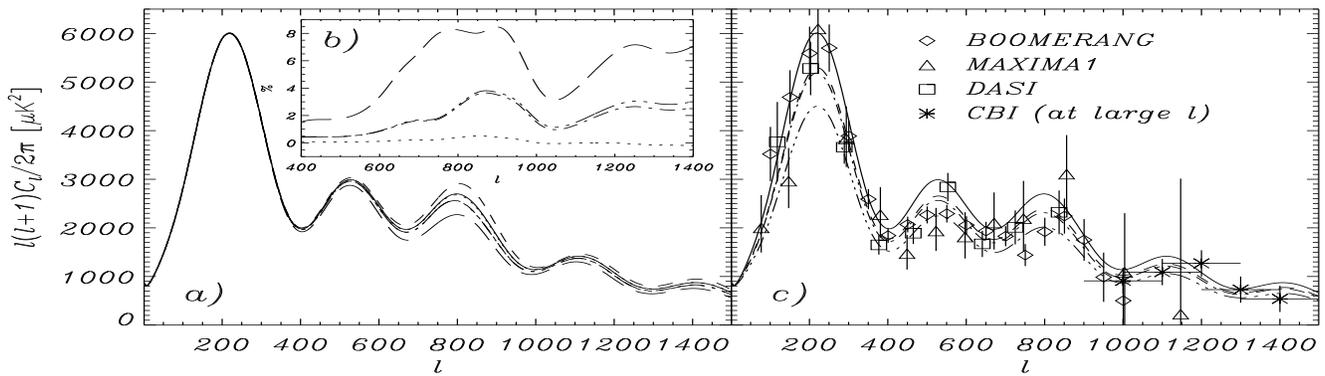,width=19.5cm,height=14cm}
\vspace{-9.8cm}
\caption{Panel a): the CMB anisotropy power spectra
(only adiabatic density mode) including the non-linear gravitational
clustering. Different values of $\Delta \alpha_{\nu}=1-\alpha_{\nu}$
are considered:
$\Delta \alpha_{\nu}= 0$ (solid line),
$\Delta \alpha_{\nu}= 2.5 \times 10^{-6}$ (dots),
$2.5\times 10^{-5}$ (dash-dots),
$1.8\times 10^{-4}$ (dash-three dots), $1\times 10^{-3}$ (long dashes).
We report also 
the case in which the non-linear gravitational
clustering and the VEP are not included (dashed line).
Panel b): relative 
differences,
in terms of 
$[1-C_l^{1/2}(\Delta \alpha_{\nu}>0)]/C_l^{1/2}(\Delta \alpha_{\nu}=0)$,
between the case $\Delta \alpha_{\nu}= 0$ and 
the cases with $\Delta \alpha_{\nu}>0$ identified 
by the same lines as in panel a).
Panel c): the CMB anisotropy power spectra 
of the four considered fiducial models compared to recent 
CMB anisotropy data: 
FM$_1$ (solid line), FM$_2$ (dashes), FM$_3$ (dash-dots),
and FM$_4$ (dash-three dots).
Power spectra normalized to COBE-{\it DMR} 4-year data \cite{Bun}
(see also the text).}
\end{figure*}
\begin{table}
\caption{$1-\sigma$ errors on the cosmological
parameters, amplitudes of different modes and  VEP
parameter $\Delta \alpha_{\nu}$ potentially
measurable with {\sc Planck} (see also the text).}
\begin{ruledtabular}
\begin{tabular}{ccccc}
 &FM$_1$ & FM$_2$ &FM$_3$ & FM$_4$\\ \hline
$n^{adi}$            & 0.00182& 0.00349 & 0.00242 & 0.0153 \\
$\tau$           & 0.0352& 0.149 & 0.217 & 0.267 \\
$\Omega_bh^2$    & 0.000175& 0.000231 & 0.000217 & 0.000327\\
$\Omega_ch^2$    & 0.00285& 0.00361 & 0.00407 & 0.00560\\
$\Omega_{\nu}h^2$&0.00155& 0.00199 & 0.00232 & 0.00327\\
$\Omega_m$       & 0.0119& 0.0152 & 0.0172 & 0.0239\\
$\Omega_{\Lambda}$& 0.00293& 0.00361 & 0.00398 & 0.00521\\
$h$               & 0.00460& 0.00582 & 0.00656 & 0.00903\\
$n^{iso}_{\nu}$&-&0.0558&-&0.285\\
$n^{iso}_{bar}$&-& -& 0.264&0.368\\
$<I_{\nu}\,I_{\nu}>$ &-   &  0.244 & - & 1.58\\
$<A \, I_{\nu}>$  &     - &  0.268 & - & 5.27\\
$<I_b\, I_b>$    &    -&      -      & 0.213& 5.21\\
$<A \, I_b> $     &    -&      -     & 0.346& 0.570\\
$<I_{\nu} \,I_b>$     &    -&      -     & -&0.368\\
$\Delta \alpha_{\nu} \times 10^{-5}$&0.936& 1.86 & 2.22&2.86
\end{tabular}
\end{ruledtabular}
\end{table}

In order to evaluate to what extent the  
{\sc Planck} surveyor can test the VEP 
by the massive neutrino component, we compute the $1-\sigma$ errors on the
estimates of the cosmological parameters and on the parameter
$\Delta \alpha_{\nu}$ by using the Fisher information
matrix approach. We consider for this computation only the {\sc Planck}
``cosmological''
channel between 70 and 217~GHz, a sky coverage $f_{sky}=0.8$
and neglect for simplicity the
foreground contamination \cite{Popa2001}.

The CMB anisotropies are sensitive to the cosmological
parameters through the evolution of 
the cosmological perturbations 
since the end of the inflation.
The simplest one-field inflation model
predicts primordial adiabatic density fluctuations. 
Isocurvature density perturbations 
can also arise in multiple field 
inflation models \cite{Enk, Linde, Langlois, Turok}. 
In order to test the sensitivity of the CMB anisotropy to 
the VEP parameter $\Delta \alpha_{\nu}$ for conditions 
more general than the pure adiabatic case,
we include in our Fisher analysis the parameters for the amplitude of the 
baryon isocurvature density mode $<I_b \, I_b> $, neutrino isocurvature
density mode $<I_{\nu} \, I_{\nu}>$, their cross-correlation
$<I_{\nu} \, I_b> $ as well as their
cross-correlations with the adibatic mode $<A \, I_b> $ and
$<A \, I_{\nu}> $, defining the following
parametric fiducial power spectrum \cite{Turok1}:
\begin{eqnarray}
C_l({\vec p})&=&C_l^A({\vec p}) \\  
&+& < I_{\nu} \, I_{\nu}> C_l^{I_{\nu}}({\vec p})+
<I_b \, I_b>C_l^{I_b}({\vec p}) \nonumber \\ 
&+& <A \, I_{\nu}>C_l^{A,I_{\nu}}({\vec p}) 
+  <A \, I_b>C_l^{A,I_b}({\vec p}) \nonumber \\               
&+& <I_{\nu} \, I_b>C_l^{I_{\nu},I_b}({\vec p}) \, ,\nonumber
\end{eqnarray}
where ${\vec p}$ represents the cosmological parameters of our
$\Lambda$CHDM cosmological model, 
$C_l^A({\vec p})$, $C_l^{I_{\nu}}({\vec p})$,
$C_l^{I_b}({\vec p})$ are the power spectra for the  adiabatic,
neutrino isocurvature and baryon isocurvature density mode,
$C_l^{A,I_{\nu}}$ and $C_l^{A,I_b}$ are the power spectra of the
cross-correlation between the adiabatic and neutrino isocurvature density
mode and adiabatic and baryon isocurvature density mode
and $C_l^{I_{\nu},I_b}({\vec p})$ corresponds to the cross-correlation
between neutrino isocurvature  and baryon isocurvature
density mode. 
The power spectra
$C_l^A$, $C_l^{I_{\nu},I_{\nu}}$ and $C_l^{I_b,I_b}$
are computed by using the CMBFAST
code \cite{Zal96} with primordial conditions including also the neutrino 
isocurvature density mode  given by \cite{Turok}. 
The cross-correlation power spectra are computed
by running the CMBFAST code with two modes excited and
then substracting the power spectra computed when each mode
was individually excited, following the prescriptions of~\cite{Turok1}.
For each mode we assume a primordial power spectrum 
with a scalar spectral index $n^{adi}=n^{iso}_{\nu}=n^{iso}_b=0.98$.
For this computation we do not 
consider the cold dark matter
isocurvature density mode, as its power spectrum 
is very similar to that obtained for
the baryon isocurvature density mode. 
We consider four fiducial models:
only the adiabatic mode (FM$_1$), 
the adiabatic mode,
the neutrino isocurvature density mode and their cross-correlation (FM$_2$), 
the adiabatic mode, the baryon isocurvature density  mode and 
their cross-correlation (FM$_3$),
and the adiabatic and all isocurvature modes as well as their
cross-correlations (FM$_4$).
Panel~c) of Fig.~1 presents the four fiducial power spectra
obtained  for $<I_b \, I_b>=0.1$, $<I_{\nu} \, I_{\nu}>=0.1 $,  
$<A \, I_b>=0.07$, $<A \, I_{\nu}>=0.07$, $<I_{\nu} \, I_b>=0.05$
and assuming $\Delta \alpha_{\nu}$=0:
they turn to be in a rough agreement with the current CMB anisotropy 
data [37--40].

Table~1 presents the $1-\sigma$ errors on the cosmological
parameters, amplitudes of different modes   
and the VEP parameter $\Delta \alpha_{\nu}$ potentially measurable 
by using the {\sc Planck} observations of the CMB temperature anisotropy.
Although the Fisher matrix approach is sensitive to
the considered fiducial model and to
the number of parameters involved in
the computation, this results shows 
that the {\sc Planck} surveyor 
will be able to place an upper limit on the VEP 
parameter $\Delta \alpha_{\nu}$ of ${\rm few} \times 10^{-5}$, valid in
the general relativity, for the case in which the gravity is the
only source of VEP. 
\begin{acknowledgments}
L.A.P. acknowledge for the financial support from
the European Space Agency. We acknowledge the 
Italy-Romania Bilateral Agreement (Project~n.~37) 
financial support. 
\end{acknowledgments}

\end{document}